\documentclass[a4paper]{llncs}
\usepackage{amsfonts}
\usepackage{tikz}

\usepackage{amsmath}
\usepackage{xcolor}
\usepackage{cite}

\begin{document}
	
	\title{veriFIRE: an Industrial Case Study in Verifying Consistency Properties
          for a DNN-Based Wildfire Detection System}
	
	\author{
		Idan Refaeli \inst{1} \and
		Maya Swisa \inst{1} \and
		Itay Buchnik \inst{2} \and
		Alon Zada \inst{2} \and
		Guy Amir \inst{3} \and
		Elad Mandelbaum\inst{2} \and
		Ziv Freund\inst{2} \and
		Guy Katz\inst{1}
	}
	\institute{
        The Hebrew University of Jerusalem, Jerusalem, Israel\\
        \email{idan.refaeli@mail.huji.ac.il, maya.swisa@mail.huji.ac.il, g.katz@mail.huji.ac.il}
          \\
          \and
        Elbit Systems - ISTAR \& EW - Elisra L.T.D, Holon, Israel\\
       \email{itay.buchnik@elbitsystems.com, alon.zada@elbitsystems.com, elad.mandelbaum@elbitsystems.com, ziv.freund@elbitsystems.com}
           \\
           \and
        Cornell University, Ithaca, New York, USA \\
        \email{gda42@cornell.edu}
	}
	
	\maketitle
	
	\begin{abstract}
    
          We present our ongoing work on the veriFIRE project: a
          collaboration between industry and academia, aimed at
          applying verification to increase the reliability of a
          real-world, safety-critical system. Specifically, we target
          an airborne platform for wildfire detection, which
          incorporates two deep neural networks.
          We present an end-to-end
          methodology for verifying \textit{consistency properties} in this
          system. Our
          approach encodes application-grounded requirements into
          solver-compatible queries for existing neural network
          verifiers. We study properties of interest over critical
          operational scenarios: (i) monotonicity of detector
          confidence as target intensity increases; and (ii) bounded
          detector response under physically plausible blur over the
          sensor. We instantiate these encodings using
          state-of-the-art neural network verification backends and
          evaluate them at scale on real background samples. For the
          first property, all verification queries are solved in under
          five minutes. For the second property, verification is
          substantially harder, highlighting key scalability
          challenges for richer, higher-dimensional
          specifications. Overall, the results demonstrate that
          meaningful, domain-specific guarantees can be obtained for
          industrial systems.
\end{abstract}

\section{Introduction}
\label{sec:introduction}

Deep neural networks (DNNs) are increasingly deployed in
safety-critical autonomous systems~\cite{KaBaDiJuKo17}. Since failures
in such settings can have severe consequences, it is important to
formally verify that a network satisfies desired behavioral
properties, rather than relying solely on
testing~\cite{KaHuIbJuLaLiShThWuZeDiKoBa19}. To support this goal, a
range of verification tools have been developed (e.g.,
Marabou~\cite{KaHuIbJuLaLiShThWuZeDiKoBa19} and
$\alpha,\beta$-CROWN~\cite{ZhWeChHsDa18,XuZhWaWaJaLiHs21,WaZhXuLiJaHsKo21}),
demonstrating strong performance on standard benchmarks
and the ability to scale to networks of practical size.

A central challenge in applying formal verification to learning-based
systems is defining the properties to be verified. Unlike traditional
software, whose behavior is specified explicitly, DNNs are learned
from data and therefore lack a direct formal specification of their
intended input-output behavior. As a result, the verification
properties must be designed separately so as to capture meaningful and
operationally relevant requirements of the system under
consideration. Even seemingly natural expectations, such as monotonic
responses to stronger signals, may be violated in practice, leading to
counter-intuitive and potentially unsafe
behavior~\cite{DuJhSaTi18}. Identifying and formalizing such
requirements is therefore an essential step toward certifying
learning-based systems~\cite{HuKwWaWu17,GeMiDrTsChVe18}.

A particular safety-critical application of DNNs is within wildfire
detection systems, whose goal is to detect and alert first responders
to situations that could later become life threatening. One such
airborne system, which is currently being considered by Elbit Systems
for use on aerial vehicles, is based on Infra-Red (IR) sensors. The IR
sensors feed their inputs --- usually a series of image frames --- to
a sequence of neural networks, which then determine whether the images
contain a wildfire.

Designing such a system to rely on DNNs affords great benefits, due to
these DNNs' ability to analyze images; but it also poses risks.
Specifically, it is possible that (a) the system will mistakenly issue
an alert when a wildfire does not exist, or, worse, that (b) the
system will fail to issue an alert when the images do indicate the
existence of a wildfire. The second kind of failure is clearly very
dangerous and could potentially jeopardize human lives. Consequently,
potential users of the system require it to be extremely
reliable. Although DNN-based systems are highly successful, prior
research has shown that even complex and highly-accurate DNNs are
prone to errors. For example, small input perturbations, due to either
random noise or adversarial attacks, are known to cause modern DNNs to
fail catastrophically~\cite{SzZaSuBrErGoFe13}. Such issues raise serious
concerns regarding the trustworthiness of a DNN-based wildfire
detection system, and could delay or prevent its deployment. We seek
to begin addressing this challenge through the application of formal
verification.

The properties we investigate are naturally derived from specific operational scenarios in which failures carry a disproportionate impact. One such scenario arises when varying the \emph{intensity} of a fire-like signal: as the signal becomes stronger, the model's confidence should not decrease. A second arises under blur variations caused by sensor manufacturing imperfections, and due to the sensor movement.

The veriFIRE project~\cite{AmFrKaMaRe23} is a collaboration between
Elbit Systems and the Hebrew University. It was initiated to address
these challenges by introducing, and verifying, consistency properties
tailored to application-specific properties of interest. In a 2022
work-in-progress paper~\cite{AmFrKaMaRe23}, these properties were
formalized at a conceptual level, capturing domain-specific
requirements such as monotonicity with respect to signal
intensity. However, such properties are naturally defined over
structured transformations of the input, and are not directly
expressible within standard verification frameworks, which operate
over fixed-input networks with bounded perturbations. In the present
work, we bridge this gap by encoding these properties of interest in a
form compatible with existing verification tools.

In this paper, we describe veriFIRE's progress since the previous report.
We present a methodology for encoding critical-section consistency
properties as formal verification queries for industrial-scale
DNNs. We instantiate this approach on two central components of a
real-world wildfire detection system: (i) the main detection network,
for which we verify monotonicity with respect to target signal
intensity; and (ii) multi-parameter verification, using a blur-related
sub-network, for which we verify that the main detection network
output is above the detection threshold for any possible blur spread
and intensity that can be generated by a specified range of
conditions. Our results demonstrate that meaningful domain-specific
guarantees over relevant scenarios can be established in practice for
industrial systems, providing a concrete step toward certifiable
learning-based sensing pipelines.

The remainder of this paper is organized as follows.
Section~\ref{sec:relatedWork} reviews related work.
Section~\ref{sec:problem} introduces the problem setting, notation,
and the properties considered in this work.
Section~\ref{sec:methodolodgy} presents our query-design methodology.
Section~\ref{sec:experimnents} reports the experimental results, and
Section~\ref{sec:discussion} discusses their implications.  Finally,
Section~\ref{sec:conclusion} concludes.

\section{Related Work}
\label{sec:relatedWork}

Formal verification for deep neural networks has progressed
substantially over the past decade. Early complete methods such as
Reluplex~\cite{KaBaDiJuKo17} demonstrated the feasibility of exact
reasoning for piecewise-linear networks, and later systems such as
Marabou~\cite{KaHuIbJuLaLiShThWuZeDiKoBa19}
extended this line into a broader verification and analysis
framework. In parallel, scalable bound-propagation-based tools, such
as
$\alpha$-$\beta$-CROWN~\cite{ZhWeChHsDa18,XuZhWaWaJaLiHs21,WaZhXuLiJaHsKo21},
have become central for handling larger models and benchmark suites.

Modern solvers incorporate a variety of techniques. These include
dedicated piecewise-linear decision procedures~\cite{Eh17},
MILP-based verification pipelines~\cite{TjXiTe17}, branch-and-bound
approaches for ReLU networks~\cite{BuLuTuToKoKu20},
abstract-interpretation and range-analysis
methods~\cite{GeMiDrTsChVe18}, abstract-domain certification
approaches~\cite{SiGePuVe19}, and output-range
reasoning~\cite{DuJhSaTi18}. Together, these techniques offer a spectrum of
trade-offs between completeness guarantees and practical runtime.  In
addition, parallelization techniques~\cite{WuOzZeIrJuGoFoKaPaBa20} are
increasingly important in industrial settings, where large query
batches must be solved under strict compute budgets. Within this
broader landscape, additional advances have focused on
abstraction-based verification, pruning and slicing
for network simplification~\cite{BuLuTuToKoKu20, WuZeKaBa22}, residual
reasoning~\cite{ElCoKa22}, and proof production~\cite{IsBaZhKa22}.

Beyond solver technology, the literature emphasizes the challenge of
specifying meaningful properties for learned systems. Foundational
work on adversarial behavior~\cite{SzZaSuBrErGoFe13} has led the
verification community to focus on robustness
properties~\cite{HuKwWaWu17}, highlighting the gap between empirical
accuracy and provable guarantees. More recent work has broadened the
scope of verification goals beyond point-wise robustness, including
verification in control settings (open-loop and
reactive systems)~\cite{DuChSa19}, explainability-oriented
checks~\cite{BaKa22}, and NLP semantics
constraints~\cite{JiRaGoLi19}. Application-grounded formulations of
such properties have also been explored in case
studies, including prior case studies on
learned robotic navigation systems~\cite{AmCoYeMaHaFaKa23}.

Our work builds on these existing results, by focusing on
application-grounded property design and query encoding for an
industrial wildfire-detection pipeline. Rather than limiting attention
to generic, norm-bounded perturbations, we target properties of
interest induced by physically meaningful transformations (target
intensity and blur), and encode them into verification queries
compatible with existing engines. In this sense, our study extends the
prior veriFIRE report~\cite{AmFrKaMaRe23} with a systematic,
deployment-scale evaluation.

\begin{figure}[ht]
\centering\includegraphics[width=0.8\columnwidth, alt={Block diagram of a two-stage airborne wildfire-detection pipeline. On the left, a drone feeds an IR image stream into a Detection network (DNN1), which outputs two candidate cropped image patches: Candidate Crop Frame (t-T) and Candidate Crop Frame (t). These two patches enter a Classification network (DNN2), depicted as a fully-connected neural network, which outputs a scalar Classification Score compared against a decision threshold tau; the alert is issued when the score exceeds tau.}]{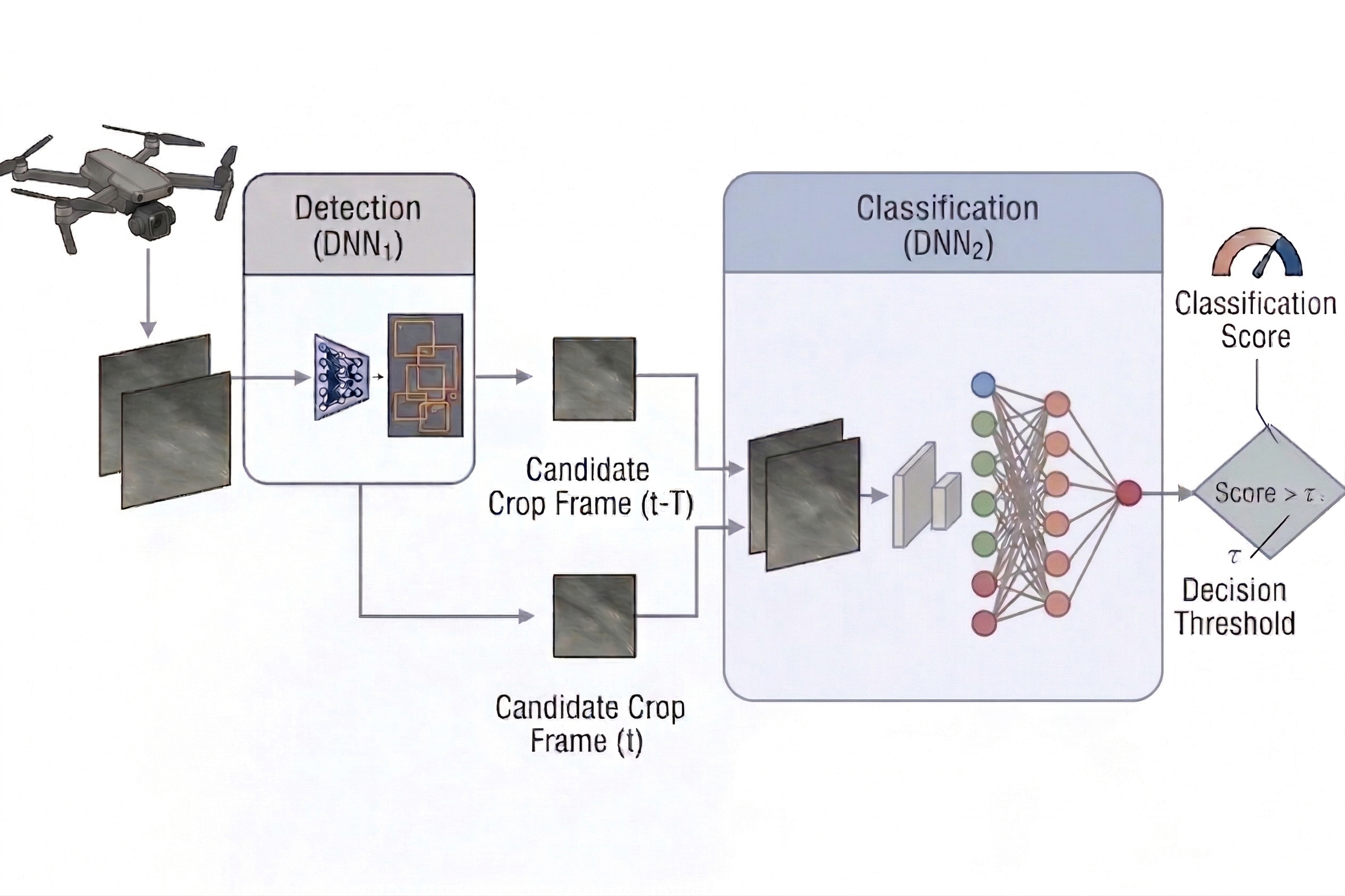}
\caption{An overview of the airborne wildfire-detection pipeline. An IR image stream is first processed by a detection network that proposes candidate regions. Each candidate is then passed to a second-stage classification network, which receives two temporally consecutive cropped frames, and outputs a confidence score. A wildfire alert is issued when the classification score exceeds the decision threshold $\tau$.}\label{fig:airborne_wildfire_detection_pipeline}
\end{figure}

\newcommand{\guy}[1]{  \marginpar{\raggedright\tiny\color{red!80!black}#1}%
}

\section{Problem Definition}
\label{sec:problem}

We consider an airborne wildfire-detection pipeline composed of two
neural networks.  The first network processes IR image streams and
proposes \emph{candidate detections}, namely cropped images containing
suspicious regions.  The second network, which is the focus of this work, receives a cropped candidate image consisting of two frames: the first is a background image from a previous frame that does not contain a fire, and the second is the current, suspected frame. The previous frame is aligned to the current frame using registration and interpolation. The network outputs a confidence score that indicates whether the current frame contains a true target. 
The goal of the network is to classify newly emerging fires under relatively low signal‑to‑noise ratio (SNR) conditions, where the fire signal is only slightly stronger than the background clutter and sensor noise. An illustration of this pipeline appears in
Figure~\ref{fig:airborne_wildfire_detection_pipeline}.  A detection
is declared whenever the score exceeds a decision threshold~$\tau$.

Training and evaluation are based on a simulator that generates synthetic, target-present samples, by planting normalized target signals into wildfire-free backgrounds of real-world records. These background records comprise extensive IR flight logs captured under varying environmental conditions to ensure robust real-world coverage. Crucially, the evaluated DNNs are already integrated and deployed within the practical, operational system. The records are two time-consecutive IR frames, previously recorded by an airborne vehicle. The target signal is then injected exclusively into the second frame, whereas the first frame remains target-free. In practice, if a fire is already weakly or partially present in the previous frame, a dedicated preprocessing algorithm handles the baseline estimation to mitigate its effect, ensuring robust detection. Consequently, the
onset of the fire corresponds to the first appearance of a
target-related signal in the second channel, allowing the classifier
to detect it immediately.

Let $B$ denote the set of admissible background crops, and let $T$
denote the set of normalized target patterns.  For
$b = \bigl(b^{(1)}, b^{(2)}\bigr) \in B$ and $t \in T$, we define
\[
x(\alpha; b,t) = \bigl(b^{(1)},\, b^{(2)} + \alpha t\bigr),
\]
where $\alpha \geq 1$ is a target-intensity factor.
Thus, $x(1;b,t)$ is the baseline planted sample, and increasing $\alpha$
strengthens the target signal while keeping the background fixed.
We denote the second network by
\[
f : \mathbb{R}^{H \times W \times 2} \to \mathbb{R},
\]
and interpret $f(x)$ as the target-confidence score of crop $x$.

\subsection{Monotonicity with Respect to Target Intensity}

Our first desired property is \emph{monotonicity}: if the same candidate crop contains
a stronger target signature, then the confidence assigned by the classifier should not decrease.
Formally, for a fixed background-target pair $(b,t)$, we require that
\[
\forall \alpha_1,\alpha_2 \in [1,\alpha_{\max}] \qquad
\alpha_1 \geq \alpha_2 \;\Longrightarrow\; f\bigl(x(\alpha_1;b,t)\bigr) \geq f\bigl(x(\alpha_2;b,t)\bigr).
\]

In our verification setup, we use an anchored version of this property relative to the baseline intensity $\alpha=1$.
Let
\[
x_{\mathrm{base}} = x(1;b,t).
\]
Then the property becomes
\[
\forall \alpha \in [1,\alpha_{\max}] \qquad
f\bigl(x(\alpha;b,t)\bigr) \geq f(x_{\mathrm{base}}).
\]

\begin{figure}[ht]
\centering\includegraphics[width=0.8\columnwidth, alt={Diagram illustrating the monotonicity property. On the left, two pairs of grayscale image patches show Channel 1 and Channel 2: the top row is the baseline target x(1;b,t) where Channel 2 contains a faint bright spot, and the bottom row is the stronger target x(alpha;b,t) with alpha greater than 1, where Channel 2 shows a brighter spot; Channel 1 is identical in both rows. A brace labels the combined input as p_ind(at,b) = (b^(1), b^(2) + at), with a note that the target is injected in the second channel only. Both inputs feed into the classifier f(·), depicted as a fully-connected network. On the right, a score plot shows two points on an increasing line: f(x(1;b,t)) at alpha=1 and f(x(alpha;b,t)) at alpha greater than 1, illustrating that the score does not decrease. A green checkbox confirms the property: f(x(alpha;b,t)) >= f(x(1;b,t)), labeled increasing target intensity does not decrease the score.}]{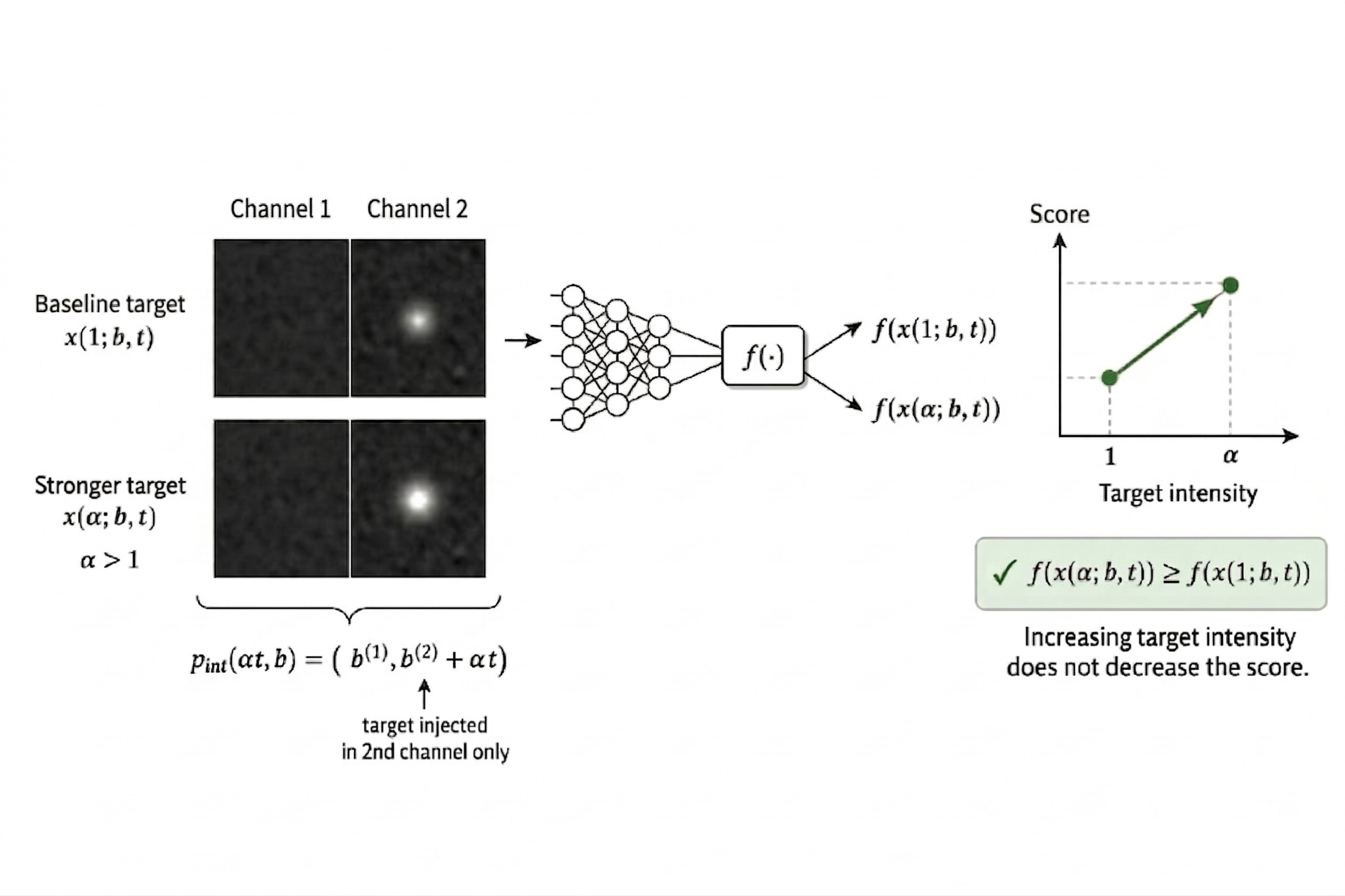}
\caption{An illustration of the monotonicity property with respect to target intensity. For a fixed background crop and target pattern, the target is injected only into the second channel, and its intensity is scaled by a factor $\alpha \geq 1$. The desired property is that increasing the target intensity should not decrease the detector confidence, i.e., $f(x(\alpha;b,t)) \geq f(x(1;b,t))$ for all admissible $\alpha$.}\label{fig:Monotonicity_pipeline}
\end{figure}

Intuitively, once a target is present in a given crop, increasing its
intensity should not reduce the detector's confidence. An illustration of
this property is described in Figure~\ref{fig:Monotonicity_pipeline}.

\begin{definition}[Local Intensity Consistency]
  Let \(f : \mathbb{R}^{H \times W \times 2} \to \mathbb{R}\) be the
  second-stage wildfire-detection network, let
  \(b = (b^{(1)}, b^{(2)}) \in \mathbb{R}^{H \times W \times 2}\) be a
  background crop, and let \(t \in \mathbb{R}^{H \times W}\) be a
  target pattern. We say that \(f\) is \((t,b)\)-locally
  intensity-consistent if for every \(\alpha \geq 1\), it holds that
  \[
  f\!\left(x(\alpha;b,t)\right)
  \geq
  f\!\left(x(1;b,t)\right),
  \]
  where \(x(\alpha;b,t)=(b^{(1)},b^{(2)}+\alpha t)\).
\end{definition}

\subsection{Blur-Tolerant Positive Detection}

Our second desired property is \emph{blur-tolerant detection}. Unlike
monotonicity with respect to target intensity, which varies a single scalar
parameter with a clear semantic interpretation (a stronger signal should not
yield lower confidence), blur tolerance captures a richer and less intuitive
physical phenomenon.

In our operational setting, fires are detected at very early stages and from
long distances. Consequently, a fire occupies a sub-pixel or near point-like
region within the sensor's instantaneous field of view (IFOV). The apparent
spatial distribution of such a point target on the sensor is therefore
dominated by the sensor's optical blur, rather than by the physical size of the
fire itself. This blur spreads the target energy over a small neighborhood of
pixels according to the sensor's optical response function.

Due to production tolerances and optical assembly variations, different sensors
--- and even different units of the same sensor model --- exhibit slightly
different blur characteristics. These variations affect how target energy is
distributed spatially across neighboring pixels, even when the physical target
and acquisition conditions are identical. As a result, the same fire may appear
with different spatial extents, local intensity profiles, and peak responses
solely due to sensor-level variability.

The blur function is supplied by the sensor manufacturer. Effectively, the blur
spreads the target energy only within a small spatial neighborhood around the
target location; in our use case, this neighborhood is limited to a
$5\times5$ region, with negligible energy outside this area. The blur behavior
is modeled by a function
\[
\psi(x,y,\sigma),
\]
where $(x,y)$ denotes the sub-pixel target location, and $\sigma$ is a
manufacturing-dependent blur parameter that controls the effective spread of
the target energy, i.e., how much the target is smeared over neighboring
pixels. For a given $(x,y,\sigma)$, this function outputs a 25-dimensional vector
corresponding to the energy distribution over a $5\times5$ pixel patch centered
at the nearest integer pixel to $(x,y)$. This mapping is provided by the sensor
manufacturer based on optical modeling and has been validated through
laboratory measurements, which confirm that its modeling error is negligible
relative to the sensor noise (NEI). The function supplied by the sensor
manufacturer is a simulation-based nonlinear function.

To formalize admissible conditions, we define the bounded intervals $X=[x_{min},x_{max}]\times[y_{min},y_{max}]$ and $\Sigma=[\sigma_{min},\sigma_{max}]$. Here, $(x,y)$ represents the sub-pixel location within the center pixel of a $5\times5$ candidate patch, which is cropped around a suspicious detection by the candidate algorithm. These offsets dictate how the intensity of the center pixel spreads to its immediate neighbors, and we incorporate target intensity $I$ explicitly by defining the intensity-augmented blur map $\psi_{I}(x,y,\sigma,I)=I\cdot\psi(x,y,\sigma)$.

Our requirement is that for any target location $(x,y)\in X$, any physically
plausible blur realization $\sigma\in\Sigma$, and any true-target intensity
within the operational range, detection should still be preserved.

\begin{figure}[ht]
\centering\includegraphics[width=0.9\columnwidth, alt={End-to-end block diagram of the blur verification pipeline, enclosed in a dashed rounded rectangle. On the left, a 4-dimensional input vector [x, y, sigma, I] enters a Blur Generation module, which outputs a small Generated Blur patch. Separately, a drone feeds an IR Image Stream of multi-channel image frames. In the central Pre Processing block, the generated blur patch undergoes Zero Padding to produce a full-size frame with a bright spot, which is added to the second channel of the IR image stream; the two channels are then concatenated via a concat operation. The resulting two-channel input enters the Classification network (DNN2), shown as a fully-connected neural network. On the right, the network outputs a Classification Score displayed on a gauge dial, which is compared against a Decision Threshold tau; the alert condition Score > tau is shown as a diamond decision node.}]{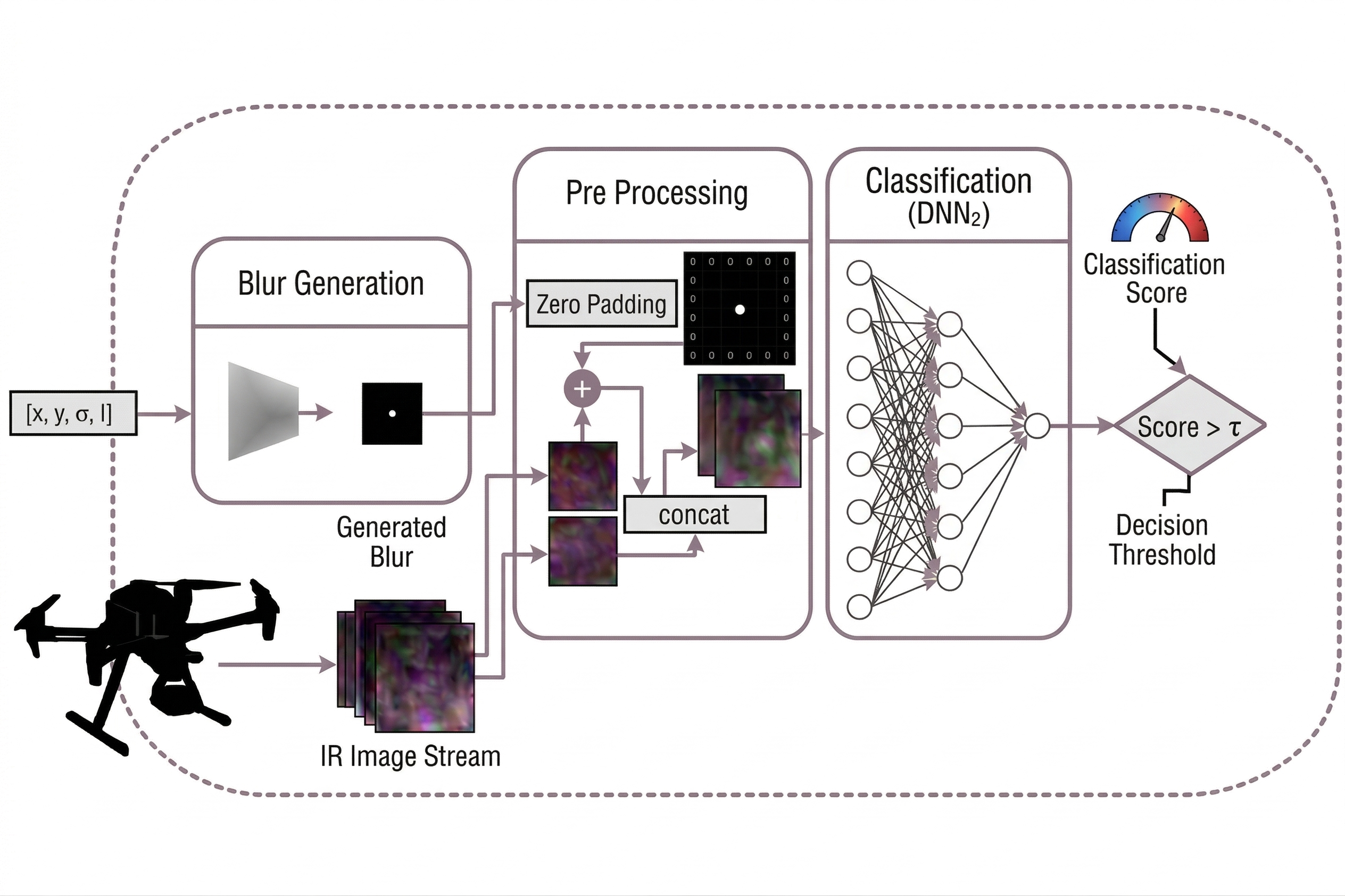}
\caption{An illustration of the end-to-end verification pipeline for blur-tolerant positive detection. A 4-dimensional parameter vector $p=(x,y,\sigma,I)$ is processed by the blur-generation module to produce a $5\times5$ target patch, which is then inserted into the second channel of the background crop and evaluated by the classifier. The verification property requires the classification score to remain above the decision threshold $\tau$ for all admissible configurations in $X\times\Sigma\times[I_0,I_1]$.}\label{fig:blur}
\end{figure}

Recall that the objective of the classification algorithm is to detect fires at
early stages while maintaining a very low false-alarm rate. We evaluated the
algorithm on extensive background flight recordings and measured the statistical
properties of the classification score for background-only detections. Given an
operational requirement on the average false-alarm rate, we set a detection
threshold $\tau$ that satisfies this requirement. In addition, we evaluated the
algorithm using traditional Monte-Carlo methods and identified a satisfactory
minimal target intensity $I_0$ that is correctly classified by the algorithm
(receives a score $>\tau$). We also define an upper operational intensity bound
$I_1$, based on sensor/manufacturing constraints. An illustration of
this property is described in Figure~\ref{fig:blur}.

We now seek to formally verify that this detection property remains consistent
across all admissible target locations, manufacturing-induced blur variations,
and target intensities in the range $[I_0,I_1]$.

Let \(b=(b^{(1)},b^{(2)})\) be a background crop, and let
\[
E_b:\mathbb{R}^{5\times5}\to\mathbb{R}^{H\times W\times 2}
\]
denote the insertion map that plants a generated blur patch into the second
channel while leaving the first channel unchanged.

Formally, for a given detection threshold \(\tau\) and minimal target intensity
\(I_0\), we require:
\[
\forall (x,y)\in X,\; \forall \sigma\in \Sigma,\; \forall I\in[I_0,I_1],\;
f\!\left(E_b\!\left(\psi_I(x,y,\sigma,I)\right)\right)\geq\tau.
\]

\section{Query Design Methodology}
\label{sec:methodolodgy}

\subsection{Monotonicity Query Design}
\label{sec:monotonicityQuery}

We reformulate each instance into a
verifier-compatible query by augmenting the network and expressing the
requirement as a standard output constraint. Modern verifiers are typically
designed for specifications that can be written as linear bounds over network
inputs/outputs, together with networks composed of common DNN operations such as
affine layers, convolutions, ReLUs, and pooling. Our consistency requirements
are defined over structured input transformations (e.g., intensity scaling and
blur generation), so they are not directly available as native property types;
we therefore encode them using only those supported operations and output
constraints.

Let $D_1=\{(b_i,t_i)\}_{i=1}^{|D_1|}$ be a dataset of background-target pairs for
the wildfire-detection network $f$. For each pair $(b_i,t_i)\in D_1$, we construct a
derived network $f_i$ by prepending a linear layer to $f$. This new layer takes
a scalar input $s\in\mathbb{R}$ (the target intensity) and outputs an image in
$\mathbb{R}^{H\times W\times 2}$.

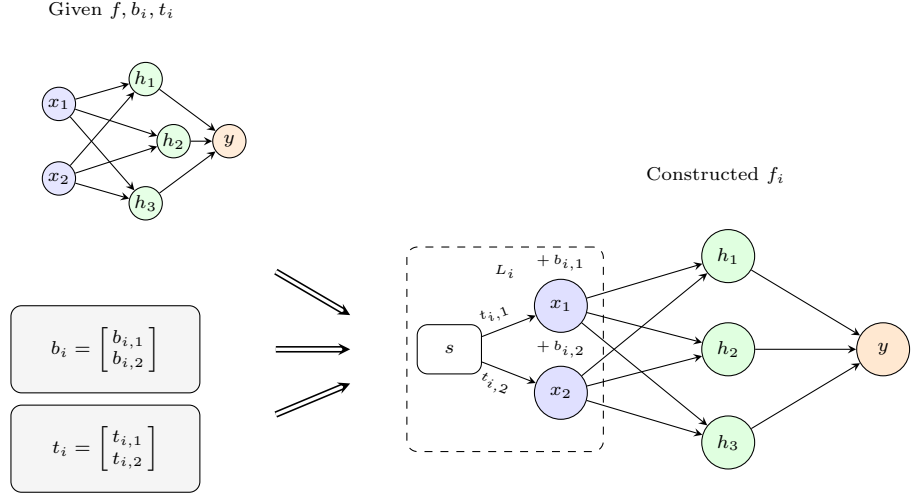
\begin{figure}[ht]
    \centering
    \begin{tikzpicture}[
        x=0.82cm,
        y=0.82cm,
        >=stealth,
        every node/.style={font=\scriptsize},
        imgstyle/.style={draw, rounded corners, align=center, minimum width=2.5cm, minimum height=1.15cm, fill=gray!8},
        instyle/.style={circle, draw, minimum size=4.4mm, inner sep=0pt, fill=blue!10},
        hidstyle/.style={circle, draw, minimum size=4.4mm, inner sep=0pt, fill=green!10},
        outstyle/.style={circle, draw, minimum size=4.4mm, inner sep=0pt, fill=orange!15},
        rinstyle/.style={circle, draw=black, minimum size=7.0mm, inner sep=0pt, fill=blue!12},
        rhstyle/.style={circle, draw=black, minimum size=7.0mm, inner sep=0pt, fill=green!12},
        routstyle/.style={circle, draw=black, minimum size=7.0mm, inner sep=0pt, fill=orange!18}
    ]
        \node[font=\scriptsize] at (-2.9,3.45) {Given $f,b_i,t_i$};

        \node[instyle] (lx1) at (-3.75,1.95) {$x_1$};
        \node[instyle] (lx2) at (-3.75,0.75) {$x_2$};
        \node[hidstyle] (lh1) at (-2.35,2.35) {$h_1$};
        \node[hidstyle] (lh2) at (-1.9,1.35) {$h_2$};
        \node[hidstyle] (lh3) at (-2.35,0.35) {$h_3$};
        \node[outstyle] (ly) at (-1.0,1.35) {$y$};

        \draw[->] (lx1) -- (lh1);
        \draw[->] (lx1) -- (lh2);
        \draw[->] (lx1) -- (lh3);
        \draw[->] (lx2) -- (lh1);
        \draw[->] (lx2) -- (lh2);
        \draw[->] (lx2) -- (lh3);
        \draw[->] (lh1) -- (ly);
        \draw[->] (lh2) -- (ly);
        \draw[->] (lh3) -- (ly);

        \node[imgstyle] (bg) at (-3.0,-2.0)
        {$b_i=\left[\begin{array}{c}b_{i,1}\\b_{i,2}\end{array}\right]$};
        \node[imgstyle] (tg) at (-3.0,-3.6)
        {$t_i=\left[\begin{array}{c}t_{i,1}\\t_{i,2}\end{array}\right]$};

        \node[font=\scriptsize] at (6.85,0.8) {Constructed $f_i$};

        \node[draw, rounded corners, minimum width=0.85cm, minimum height=0.65cm] (s) at (2.55,-2.0) {$s$};

        \draw[->, double, line width=0.60pt, double distance=1.25pt] (-0.25,-0.75) -- (0.95,-1.45);
        \draw[->, double, line width=0.60pt, double distance=1.25pt] (-0.25,-2.00) -- (0.95,-2.00);
        \draw[->, double, line width=0.60pt, double distance=1.25pt] (-0.25,-3.05) -- (0.95,-2.55);

        \node[draw,dashed,rounded corners,minimum width=2.6cm,minimum height=2.7cm] (lbox) at (3.45,-2.0) {};
        \node[font=\tiny] at (3.45,-0.75) {$L_i$};

        \node[rinstyle] (rx1) at (4.35,-1.3) {$x_1$};
        \node[rinstyle] (rx2) at (4.35,-2.7) {$x_2$};
        \node[font=\tiny] at (4.35,-0.60) {$+\,b_{i,1}$};
        \node[font=\tiny] at (4.35,-2.00) {$+\,b_{i,2}$};

        \draw[->] (s) -- node[pos=0.35, above, sloped, font=\tiny] {$t_{i,1}$} (rx1);
        \draw[->] (s) -- node[pos=0.35, below, sloped, font=\tiny] {$t_{i,2}$} (rx2);

        \node[rhstyle] (rh1) at (7.05,-0.5) {$h_1$};
        \node[rhstyle] (rh2) at (7.05,-2.0) {$h_2$};
        \node[rhstyle] (rh3) at (7.05,-3.5) {$h_3$};
        \node[routstyle] (ry) at (9.55,-2.0) {$y$};

        \draw[->] (rx1) -- (rh1);
        \draw[->] (rx1) -- (rh2);
        \draw[->] (rx1) -- (rh3);
        \draw[->] (rx2) -- (rh1);
        \draw[->] (rx2) -- (rh2);
        \draw[->] (rx2) -- (rh3);
        \draw[->] (rh1) -- (ry);
        \draw[->] (rh2) -- (ry);
        \draw[->] (rh3) -- (ry);
    \end{tikzpicture}
    \caption{Construction of the verification network used for monotonicity queries. Given a detector $f$, a background crop $b_i$, and a target pattern $t_i$, we prepend a scalar-input linear layer $L_i(s)=b_i+s\tilde{t}_i$, where $\tilde{t}_i$ is zero in the first channel and equals $t_i$ in the second channel. The resulting composed network $f_i(s)=f(L_i(s))$ enables the monotonicity requirement to be encoded as a standard verifier-compatible output constraint over the scalar input range $s \in [1,\alpha_{\max}]$.}
    \label{fig:fi-network-construction}
\end{figure}

Formally, the prepended layer implements
\[
L_i(s)=b_i+s\cdot\tilde{t}_i,
\]
where $\tilde{t}_i\in\mathbb{R}^{H\times W\times 2}$ is zero in the first
channel and equals $t_i$ in the second channel. Equivalently, the layer's
weights are the target pattern (second channel only) and its bias is the
background crop. Figure~\ref{fig:fi-network-construction} illustrates this
construction on a toy example, where we show a detector $f$, vectors
$b_i,t_i$, and construct $f_i$ by prepending a scalar-input linear layer with
$W=t_i$ and $b=b_i$. The composed verification network is
\[
f_i(s)=f\!\left(L_i(s)\right).
\]
The construction of $f_i$ is also illustrated in
Figure~\ref{fig:fi-network-construction}.

For each $i$, we precompute the baseline output at $s=1$,
\[
y_i^{\mathrm{base}}=f_i(1),
\]
which corresponds to the original, unscaled target.

The property constrains the input scalar to the interval
$s\in[1,\alpha_{\max}]$ and requires
\[
f_i(s)\geq y_i^{\mathrm{base}}.
\]
Equivalently,
\[
\forall s\in[1,\alpha_{\max}]:\; f_i(s)-y_i^{\mathrm{base}}\geq 0.
\]
Hence, each sample in $D_1$ yields one query that checks monotonicity for its
specific background-target pair.

\subsection{Blur Query Design}
\label{sec:blurQuery}

Unlike the monotonicity case, where a fixed linear layer is added before the
detector, the blur case verifies a composed model in which the blur-generation
network precedes the detector.

As described in Section~\ref{sec:problem}, the sensor manufacturer supplies a
nonlinear blur-response function $\psi(x,y,\sigma)$, and we extend it to an
intensity-augmented map $\psi_I(x,y,\sigma,I)=I\cdot\psi(x,y,\sigma)$. Directly
incorporating this exact simulation-based map into a verification query is hard
for current DNN verification backends, because it involves nonlinear operations
outside the standard verifier-supported network graph.

To obtain verifier-compatible queries, we emulate the manufacturer blur function
with an additional neural network. This auxiliary network is then composed with
the detector, so that the end-to-end query is expressed using supported
operations only.

The blur-generation network is trained offline to accurately approximate the
manufacturer-provided analytical blur model. Specifically, training data are
generated by sampling a four-dimensional parameter vector
$p=(x,y,\sigma,I)$, where $x$ and $y$ denote spatial offsets, $\sigma$
controls the blur spread, and $I$ represents target intensity; we then compute
the corresponding target patches using the exact nonlinear blur function
supplied by the sensor manufacturer. The network is trained in a supervised
manner to regress from the parameter vector to the resulting $5\times 5$ target
patch. Model fidelity is assessed on a held-out validation set, and both
training and validation errors are verified to lie below the sensor's
lower-level noise floor. Consequently, the approximation error introduced by
replacing the analytical blur function with the learned network is negligible
relative to the inherent sensor noise, justifying its use within the formal
verification queries.

Let $D_2=\{b_i\}_{i=1}^{|D_2|}$ be a dataset of background crops for the
wildfire-detection network $f$. In addition, let
\[
B:\mathbb{R}^4\rightarrow\mathbb{R}^{5\times 5}
\]
denote the blur-generation network. Its input is a 4-dimensional parameter
vector
\[
p=(x,y,\sigma,I),
\]
where $x$ and $y$ represent spatial offsets, $\sigma$ is the
manufacturing-dependent blur parameter, and $I$ is the target intensity. For
each such parameter vector, the network $B$ outputs a $5\times 5$ target patch
that approximates $\psi_I(x,y,\sigma,I)$.

For each background crop $b_i\in D_2$, we construct a derived network $g_i$ by
composing the blur-generation network $B$, a fixed insertion map $E_i$, and
the original detector $f$. The map $E_i$ leaves the background crop $b_i$
unchanged and incorporates the generated $5\times 5$ patch into the detector
input. The composed verification network is
\[
g_i(p)=f\!\left(E_i(B(p))\right).
\]
This composition follows the same network-augmentation pattern shown in
Figure~\ref{fig:fi-network-construction}, with a four-parameter blur generator
in place of the single-scalar monotonicity layer.
In the implementation, $E_i$ is realized as a non-trainable dense layer whose
bias is the flattened background crop $b_i$, and whose weights map the $5\times 5$
outputs of $B$ to the central $5\times 5$ region of the second channel, while
leaving all other pixels unchanged.

Let
\[
\mathcal{P}=X\times\Sigma\times[I_0,I_1]
\]
denote the admissible parameter box. The property then requires that
the detector output remain above a fixed threshold $\tau$ throughout this
region:
\[
\forall p\in\mathcal{P}:\; g_i(p)\geq \tau.
\]
More generally, if the detector output is vector-valued, the property
enforces the bound component-wise.

Hence, each background crop in $D_2$ yields one query that checks whether any
admissible blur configuration can cause the detector response to fall below the
prescribed threshold.

\section{Experimental Results}
\label{sec:experimnents}

\subsection{Experimental Setup}
\label{sec:experimentalSetup}

We report experiments for both properties using one verification query per data
sample. For each query, we record the solver status (SAT, UNSAT, or timeout)
and wall-clock solving time. We report averages over all queries, and when
applicable also stratify runtime by SAT and UNSAT outcomes.

Our methodology is backend-agnostic, and can be instantiated with any
verification engine that supports the encoded networks and property files. In
this work, we instantiate it using $\alpha,\beta$-CROWN.

All queries are executed as independent jobs on a Slurm cluster. For
reproducibility, each query is encoded as a network file together with a formal
property specification in a verifier-compatible format (ONNX + VNNLIB).

Both sets of experiments use the same benchmark collection of $2011$ IR
background crops, each of shape $25\times25\times2$. For monotonicity, each
background is paired with its corresponding target pattern,
yielding $2011$ background-target pairs. For blur, we use a subset of $1698$
background images from this same collection.

The DNN architectures used in the experiments are summarized in
Tables~\ref{tab:detector-architecture} and \ref{tab:blur-architecture}.

\begin{table}[t]
\centering
\caption{Detector architecture ($f$) used for monotonicity and blur queries.}
\label{tab:detector-architecture}
\begin{tabular}{lrr}
\hline
Layer (type) & Output shape & Params \\
\hline
Conv2D & $(23,23,64)$ & 1,216 \\
MaxPool2D & $(11,11,64)$ & 0 \\
BatchNorm & $(11,11,64)$ & 256 \\
Conv2D & $(9,9,64)$ & 36,928 \\
MaxPool2D & $(4,4,64)$ & 0 \\
BatchNorm & $(4,4,64)$ & 256 \\
Conv2D & $(2,2,64)$ & 36,928 \\
MaxPool2D & $(1,1,64)$ & 0 \\
Flatten & $(64)$ & 0 \\
Dense & $(256)$ & 16,640 \\
Dense & $(1)$ & 257 \\
\hline
Total parameters & -- & 92,481 \\
\hline
\end{tabular}
\end{table}

\begin{table}[t]
\centering
\caption{Blur-generation architecture ($B$) used in blur queries.}
\label{tab:blur-architecture}
\begin{tabular}{lrr}
\hline
Layer (type) & Output shape & Params \\
\hline
Input $(x,y,\sigma,I)$ & $(4)$ & 0 \\
fc1 (Dense) & $(64)$ & 320 \\
fc2 (Dense) & $(128)$ & 8,320 \\
fc3 (Dense) & $(64)$ & 8,256 \\
fc4 (Dense) & $(25)$ & 1,625 \\
\hline
Total parameters & -- & 18,521 \\
\hline
\end{tabular}
\end{table}

\subsubsection{Monotonicity Setup.}

For each background-target pair $(b_i,t_i)$, we generate one
verification query with intensity range $s\in[1,\alpha_{\max}]$ and set
$\alpha_{\max}=2$.

This setup yields $2011$ independent monotonicity verification tasks, each with
a scalar input variable and a fixed intensity interval. Each query is
allocated one CPU and $64$GB RAM.

\subsubsection{Blur Setup.}

For blur queries, we generate one query for each background image in the
selected subset.

This setup yields $1698$ independent blur verification tasks.
Each query is allocated one GPU and $64$GB RAM, with a
timeout of $4$ hours. Each query is defined over a four-parameter box, which is a richer search space than in the
monotonicity case. For each query, verification is performed over the following input-parameter
bounds; the exact values of \(\Sigma\), \(I_0\), and \(I_1\) cannot be shared
due to confidentiality constraints:
\[
x\in[-0.5,0.5],\quad
y\in[-0.5,0.5],\quad
\sigma\in \Sigma,\quad
I\in[I_0,I_1].
\]

\subsection{Monotonicity Queries}
\label{sec:monotonicityResults}

\subsubsection{Results.}

Table~\ref{tab:monotonicity-outcomes} reports the outcome distribution. All
$2011$ queries are solved within $5$ minutes. In our encoding, an UNSAT result
means the monotonicity property is proven for the sample over
$[1,\alpha_{\max}]$, while a SAT result indicates a counterexample.

\begin{table}[t]
\centering
\caption{Outcome statistics for monotonicity queries ($\alpha_{\max}=2$).}
\label{tab:monotonicity-outcomes}
\begin{tabular}{lrr}
\hline
Result type & Count & Percent \\
\hline
UNSAT (property holds) & 1134 & 56.39\% \\
SAT (counterexample found) & 877 & 43.61\% \\
Total & 2011 & 100.00\% \\
\hline
\end{tabular}
\end{table}

Table~\ref{tab:monotonicity-runtime} summarizes runtime statistics. The average
verification time is $225.81$ seconds overall, with SAT instances solved
faster on average than UNSAT instances.

Overall, the monotonicity property is proven for a slight majority of the
evaluated samples (UNSAT: $56.39\%$), while counterexamples are found for the
remaining cases (SAT: $43.61\%$). Runtime-wise, UNSAT instances are slower by
about $25.19$ seconds on average, which is consistent with the additional proof
effort required to certify the property over the full interval.

\begin{table}[t]
\centering
\caption{Runtime statistics for monotonicity queries (seconds).}
\label{tab:monotonicity-runtime}
\begin{tabular}{lr}
\hline
Statistic & Time (s) \\
\hline
Average over all queries & 225.810 \\
Average over SAT queries & 211.608 \\
Average over UNSAT queries & 236.793 \\
\hline
\end{tabular}
\end{table}

\subsection{Blur Queries}
\label{sec:blurResults}

\subsubsection{Results.}

Table~\ref{tab:blur-summary} reports the blur-query outcome distribution and
runtime statistics.

We run the experiments on $2011$ queries, which are based on the same set of
the background crops used in the monotonicity experiments.

The reported blur results indicate that verifying the blur property is more
challenging than verifying monotonicity. This observation motivates ongoing
work on improved architectures and verification workflows.

We summarize the key quantitative findings through three aggregates: (i) the
SAT/UNSAT/timeout balance; (ii) the relative SAT-vs.-UNSAT solving cost; and
(iii) the aggregate runtime profile under the current computational budget.

\begin{table}[t]
\centering
\caption{Blur-query outcome and runtime summary.}
\label{tab:blur-summary}
\begin{tabular}{lr}
\hline
Metric & Value \\
\hline
Number of SAT queries & 212 \\
Number of UNSAT queries & 548 \\
Number of timeout queries & 1251 \\
Average solving time (SAT) [s] & 43.50 \\
Average solving time (UNSAT) [s] & 8013.98 \\
Average solving time (all queries) [s] & 2188.41 \\
\hline
\end{tabular}
\end{table}

\section{Discussion}
\label{sec:discussion}

The results reveal a clear difference in verification difficulty
between the two properties studied in this work. The monotonicity
queries are relatively tractable: all instances are solved within the
given time budget, and a substantial portion are formally proven. In
contrast, the blur queries are significantly more challenging, with
many instances timing out. This gap suggests that verification
complexity is strongly influenced not only by the detector itself, but
also by the dimensionality and structure of the property being
verified.

The results also demonstrate the value of verification beyond standard
empirical evaluation. SAT outcomes identify concrete samples for which the desired property fails, while UNSAT outcomes provide formal guarantees for specific instances. Currently, SAT counterexamples undergo root-cause analysis to pinpoint specific background textures or target configurations causing failures, directly guiding targeted data augmentation and model retraining. Meanwhile, timeout cases primarily inform necessary adjustments to our verification workflow, such as refining domain bounds or tweaking solver heuristics. In this sense, verification offers a complementary perspective: rather than measuring average behavior,
it exposes whether the detector satisfies application-relevant consistency requirements on operational scenarios.

From an industrial perspective, these results are a valuable
milestone, although additional work is required before verification
can be applied as part of the standard development pipeline ---
particularly regarding verifier scalability, as real industrial models
are often larger than those studied here. Still, even unresolved
queries provide practical value: they expose challenging cases and
weak areas of the model, which can be fed back into the development
cycle to improve training, evaluation, and future model design. In
addition, this work provides a clearer understanding of how to
translate operational quality requirements into formal verification
queries, giving rise to a practical methodology that can support
future architectures and other learning-based systems.

\section{Conclusion and Future Work}
\label{sec:conclusion}

We introduced an end-to-end methodology for verifying
application-grounded consistency properties in an industrial
wildfire-detection pipeline. By encoding monotonicity with respect to
target intensity and positive detection under physically plausible
blur as solver-compatible queries, we showed how formal verification
can be applied to real DNN-based sensing systems.

Academically, the results demonstrate both promise and challenge: monotonicity can be verified at practical scale, with all 2011 queries solved within five minutes, while blur-based properties remain significantly harder because they involve a richer physical parameterization. Industrially, the value is twofold: verified cases provide concrete guarantees, and failed or unresolved cases expose model weaknesses that can directly inform retraining and redesign.

Future work will focus on developing a systematic pipeline to verify the
auxiliary blur-generation network, ensuring full verification completeness
across the sensing-to-detection chain. Additionally, we intend to perform a
root-cause analysis on samples that could not be formally verified --- to
identify specific input characteristics or model architectural features that
impede formal certification.

Overall, this work suggests that formal verification is already useful as both an assurance mechanism and a development tool for safety-critical ML pipelines. Future work will focus on improving scalability, strengthening architecture-verification co-design, and extending coverage to additional operational scenarios.

\subsubsection{Acknowledgements.}  
This work of Refaeli, Swisa and Katz was partially funded by the European Union
(ERC, VeriDeL, 101112713). Views and opinions expressed
are however those of the author(s) only and do not necessarily reflect those of the European Union or the European
Research Council Executive Agency. Neither the European
Union nor the granting authority can be held responsible for
them. This research was additionally supported by a grant
from the Israeli Science Foundation (grant number 558/24).

{
\bibliographystyle{abbrv}
\bibliography{references}

@inproceedings{AmFrKaMaRe23,
	Title = {{veriFIRE: Verifying an Industrial, Learning-Based Wildfire 
	Detection System}},
	Author = {Amir, G. and Freund, Z. and Katz, G. and Mandelbaum, E. and 
	Refaeli, I.},
	Booktitle = {Proc. 25th Int. Symposium on Formal Methods (FM)},
	Year = {2023},
	Pages = {648--656},
}

@inproceedings{KaBaDiJuKo17,
	title = {{Reluplex: An Efficient SMT Solver for Verifying Deep Neural 
	Networks}},
	author = {Katz, G. and Barrett, C. and Dill, D. and Julian, K. and 
	Kochenderfer, M.},
	year = {2017},
	booktitle = {Proc. 29th Int. Conf. on Computer Aided Verification (CAV)},
	pages = {97--117}
}

@inproceedings{ZhWeChHsDa18,
  title={{Efficient Neural Network Robustness Certification with General Activation Functions}},
  author={Zhang, H. and Weng, T.-W. and Chen, P.-Y. and Hsieh, C.-J. and Daniel, L.},
  Booktitle = {Proc. 32nd Conf. on Neural Information Processing Systems (NeurIPS)},
  pages={4939--4948},
  year={2018},
}

@inproceedings{XuZhWaWaJaLiHs21,
    title={{Fast and Complete: Enabling Complete Neural Network Verification with Rapid and Massively Parallel Incomplete Verifiers}},
    author={Xu, K. and Zhang, H. and Wang, S. and Wang, Y. and Jana, S. and Lin, X. and Hsieh, C.-J.},
    Booktitle = {Proc. 38th Int. Conf. on Machine Learning (ICML)},
    year={2021},
}

@inproceedings{WaZhXuLiJaHsKo21,
  title     = {{Beta-CROWN: Efficient Bound Propagation with Per-Neuron Split Constraints for Neural Network Robustness Verification}},
  author    = {Wang, Shiqi and Zhang, Huan and Xu, Kaidi and Lin, Xue and Jana, Suman and Hsieh, Cho-Jui and Kolter, Zico},
  booktitle = {Proc. 35th Conf. on Neural Information Processing Systems (NeurIPS)},
  year      = {2021}
}

@inproceedings{KaHuIbJuLaLiShThWuZeDiKoBa19,
  title     = {{The Marabou Framework for Verification and Analysis of Deep Neural Networks}},
  author    = {Katz, Guy and Huang, Derek and Ibeling, Duligur and Julian, Kyle and Lazarus, Christopher and Lim, Rachel and Shah, Parth and Thakoor, Shantanu and Wu, Haoze and Zelji{\'c}, Aleksandar and Dill, David and Kochenderfer, Mykel and Barrett, Clark},
  booktitle = {Proc. 31st Int. Conf. on Computer Aided Verification (CAV)},
  pages     = {443--452},
  year      = {2019}
}

@Misc{SzZaSuBrErGoFe13,
	Title  = {{Intriguing Properties of Neural Networks}},
	Author = {Szegedy, C. and Zaremba, W. and Sutskever, I. and Bruna, J. and 
	Erhan, D. and Goodfellow, I. and Fergus, R.},
	Note  = {Technical Report. \url{http://arxiv.org/abs/1312.6199}},
	Year  = {2013}
}

@inproceedings{GeMiDrTsChVe18,
	Title = {{AI2: Safety and Robustness Certification of Neural Networks with 
	Abstract Interpretation}},
	Author = {Gehr, T. and Mirman, M. and Drachsler-Cohen, D. and Tsankov, E. 
	and Chaudhuri, S. and Vechev, M.},
	year = {2018},
	booktitle = {Proc. 39th IEEE Symposium on Security and Privacy (S\&P)},
}

@inproceedings{HuKwWaWu17,
	title = {{Safety Verification of Deep Neural Networks}},
	author = {Huang, X. and Kwiatkowska, M. and Wang, S. and Wu, M.},
	year = {2017},
	booktitle = {Proc. 29th Int. Conf. on Computer Aided Verification (CAV)},
	pages = {3--29}
}

@inproceedings{DuJhSaTi18,
	title={{Output Range Analysis for Deep Feedforward Neural Networks}},
	author={Dutta, S. and Jha, S. and Sankaranarayanan, S. and Tiwari, A.},
	booktitle={Proc. 10th NASA Formal Methods Symposium (NFM)},
	pages={121--138},
	year={2018},
}

@inproceedings{WuOzZeIrJuGoFoKaPaBa20,
	Title = {{Parallelization Techniques for Verifying Neural Networks}},
	Author = {Wu, H. and Ozdemir, A. and Zelji\'c, A. and Irfan, A. and Julian, 
	K. and Gopinath, D. and Fouladi, S. and Katz, G. and P\u{a}s\u{a}reanu, C. 
	and Barrett, C.},
	booktitle = {Proc. 20th Int. Conf. on Formal Methods in Computer-Aided 
	Design (FMCAD)},
	Year = {2020},
	pages = {128--137}
}

@inproceedings{Eh17,
	author    = {Ehlers, R.},
	title     = {{Formal Verification of Piece-Wise Linear Feed-Forward Neural 
	Networks}},
	Booktitle = {Proc. 15th Int. Symp. on Automated Technology for Verification 
	and Analysis (ATVA)},
	pages     = {269--286},
	year      = {2017},
}

@Misc{TjXiTe17,
	Title = {{Evaluating Robustness of Neural Networks with Mixed Integer 
	Programming}},
	Author = {Tjeng, V. and Xiao, K. and Tedrake, R.},
	Note = {Technical Report. \url{http://arxiv.org/abs/1711.07356}},
	Year = {2017}
}

@inproceedings{SiGePuVe19,
    Title                    = {{An Abstract Domain for Certifying Neural Networks}},
  Author                   = {Singh, G. and Gehr, T. and Puschel, M. and Vechev, M.},
  Booktitle                = {Proc. 46th ACM SIGPLAN Symposium on Principles of Programming Languages (POPL)},
  Year                     = {2019},

}

@article{BuLuTuToKoKu20,
  title   = {{Branch and Bound for Piecewise Linear Neural Network Verification}},
  author  = {Bunel, Rudy and Lu, Jingyue and Turkaslan, Ilker and Torr, Philip H. S. and Kohli, Pushmeet and Kumar, M. Pawan},
  journal = {Journal of Machine Learning Research},
  year    = {2020},
  url     = {https://jmlr.org/papers/v21/19-468.html}
}

@inproceedings{WuZeKaBa22,
  title     = {{Efficient Neural Network Analysis with Sum-of-Infeasibilities}},
  author    = {Wu, Haoze and Zelji{\'c}, Aleksandar and Katz, Guy and Barrett, Clark},
  booktitle = {Proc. 28th Int. Conf. on Tools and Algorithms for the Construction and Analysis of Systems (TACAS)},
  pages     = {143--163},
  year      = {2022}
}

@inproceedings{ElCoKa22,
  Title = {{Neural Network Verification using Residual Reasoning}},
  Author = {Elboher, Y. and Cohen, E. and Katz, G},
  Booktitle = {Proc. 20th Int. Conf. on Software Engineering and Formal Methods
  (SEFM)},
  Year = {2022},
  Pages = {173--189},
}

@InProceedings{IsBaZhKa22,
  Title = {{Neural Network Verification with Proof Production}},
  Author = {Isac, O. and Barrett, C. and Zhang, M. and Katz, G.},
  Booktitle = {Proc. 22nd Int. Conf. on Formal Methods in Computer-Aided Design
  (FMCAD)},
  Pages = {38--48}, 
  Year = {2022}
}

@inproceedings{AmCoYeMaHaFaKa23,
	title={{Verifying Learning-Based Robotic Navigation Systems}},
	author={Amir, Guy and Corsi, Davide and Yerushalmi, Raz and Marzari, Luca and Harel, David and Farinelli, Alessandro and Katz, Guy},
	booktitle={Proc. 29th Int. Conf. on Tools and Algorithms for the Construction and Analysis of Systems (TACAS)},
	pages={607--627},
	year={2023},
}

@InProceedings{DuChSa19,
  Title = {{Reachability Analysis for Neural Feedback Systems using Regressive Polynomial Rule Inference}},
  Author = {Dutta, S. and Chen, X. and Sankaranarayanan, S.},
  Booktitle = {Proc. 22nd ACM Int. Conf. on Hybrid Systems: Computation and Control (HSCC)},
  pages = {157--168},
  Year = {2019},
 }

@inproceedings{BaKa22,
  title={{Towards Formal Approximated Minimal Explanations of Neural Networks}},
  author={Bassan, S. and Katz, G.},
  booktitle={Proc. 29th Int. Conf. on Tools and Algorithms for the Construction and Analysis of Systems (TACAS)},
  pages = {187--207},
  year={2023}
}

@inproceedings{JiRaGoLi19,
  title = {{Certified robustness to adversarial word substitutions}},
  author = {Jia, R. and Raghunathan, A. and G{\"o}ksel, K. and Liang, P.},
  booktitle = {Proc. Conf. on Empirical Methods in Natural Language Processing and 9th Int. Joint Conf. on Natural Language Processing (EMNLP-IJCNLP)},
  pages = {4129--4142},
  year = {2019},
  doi = {10.18653/v1/D19-1423}
}
}
	
\end{document}